# The interactions between antiferromagnetism, tetrahedral sites and electron-phonon coupling in FeSe$_{1-x}$Te$_x$ and FeSe/SrTiO$_3$

[1,2]C. H. Wong[*] and [1]R. Lortz[*]

[1]Department of Physics, The Hong Kong University of Science and Technology, Clear Water Bay, Kowloon, Hong Kong

[2]Institute of Physics and Technology, Ural Federal University, Russia

[*]ch.kh.vong@urfu.ru, [*]lortz@ust.hk

**Abstract**

We show that the superconducting transition temperature $T_c$ of FeSe$_{1-x}$Te$_x$ can be computed to reasonable values in a modified McMillan approach in which the electron-phonon coupling is amplified by the antiferromagnetism and the out-of-plane phonons triggered by the tetrahedral lattice sites. This interplay is not only effective at ambient pressure, but also under hydrostatic compression. According to our model, the theoretical $T_c$ of the compressed FeSe$_{0.5}$Te$_{0.5}$ agrees with experiment results. More importantly, by taking into account the interfacial effect between an FeSe monolayer and its SrTiO$_3$ substrate as an additional gain factor, our calculated $T_c$ value is up to 91 K high, and provides evidence that the strong $T_c$ enhancement recently observed in such monolayers with $T_c$ reaching 100 K may be due to an enhanced-electron phonon coupling.

**Introduction**

Iron-based superconductors feature a rich phase diagram with multiple forms of electronic order [1-7]. In addition to superconductivity, they usually exhibit an antiferromagnetic spin density wave phase on the underdoped side of the phase diagram [8], more or less coinciding with a nematic phase associated with an electronic instability [9-15] that causes a pronounced anisotropy of the Fe-As layers in the plane. This nematic phase was recently considered a vestigial order to a spin density wave order with a multi-component order parameter [16]. Although it is generally assumed that magnetism should play a major role in the superconducting coupling mechanism, the exact relation of these additional electronic phases to superconductivity is not yet fully understood. The presence of these additional electronic phases makes the theoretical analysis of superconductivity in this class of materials extremely sophisticated without a comprehensive model to explain the high transition temperatures.

Among the iron-based superconductors, FeSe has the simplest structure, consisting of sheets of two-dimensional FeSe layers stacked on top of each other without additional ions as charge reservoir between the layers. It becomes superconducting below 8 K [17]. FeSe is also simpler in the sense that it is non-magnetic and has only a nematic order, which is formed well above the superconducting transition temperature [18,19]. Under pressure, however, it features an equivalently rich phase diagram as other iron based compounds. $T_c$ is first increased to 38 K at 4 GPa [20,21], then a spin density wave phase is formed which suppresses $T_c$, while at higher pressure a re-emerging superconductivity with a maximum $T_c$ of 48 K occurs. FeSe can also be doped by partially replacing Se by Te with an optimized $T_c$ of FeSe$_{1-x}$Te$_x$ at $x = 0.5$ [21].

The layer structure of FeSe makes it possible to grow monolayers of FeSe epitaxially on a substrate. In 2013, superconductivity was reported with a record $T_c$ of 70 K on monolayer FeSe on a SrTiO$_3$ substrate [22], which was later increased to 100 K [23].

Despite the complexity of the electronic phase diagram of iron–based superconductors, which suggests the presence of additional broken symmetries besides the broken U(1) gauge symmetry of the superconducting state and thus an unconventional pairing mechanism, recent works have suggested that the role of electron-phonon coupling could play a certain role in the superconducting mechanism of iron-based superconductors [24-26], although there is clear evidence that magnetic fluctuations must be taken into account. The high transition temperature of the monolayer FeSe on a SrTiO$_3$ substrate gives further indications of the importance of electron-phonon coupling. While growing FeSe films on graphene substrate suppresses $T_c$ [27], the giant enhancement of $T_c$ is likely activated by the SrTiO$_3$ substrate, where the interfacial contribution cannot be ignored. Strong electron–phonon coupling at the interface of FeSe/SrTiO$_3$ has been identified in ARPES data [28], with electrons located 0.1-0.3eV below the Fermi level involved in superconductivity. Although the FeSe phonons do not depend on the thickness of the FeSe material, the F-K phonon across the interface may be responsible for the high $T_c$ [29]. According to the experiment by S. Zhang *et al* [29], the F-K phonons of the FeSe/SrTiO$_3$ surface show new energy loss modes and the line width is widened compared to bare SrTiO$_3$.

We have recently shown that it is possible to explicitly calculate superconducting transition temperatures of various iron-based superconductors, including LiFeAs, NaFeAs, FeSe, BaFe$_2$As$_2$ and Ba$_{1-x}$K$_x$Fe$_2$As$_2$ [30,31], using a modified Mc-Millan approach that takes into account an enhancement of the electron-phonon coupling by local antiferromagnetic order using ab-initio parameters as input. Here we use this model to test whether such an approach can be applied to the bulk FeSe$_{1-x}$Te$_x$ system and to test whether the interfacial phonon can actually explain the 100-K superconductivity in FeSe/SrTiO$_3$.

**Theory**

Our theoretical approach to iron-based superconductors is revisited here [30]. Many bulk iron-based superconductors share the same characteristic in the ARPES data, i.e. a noticeable shift of spectral weight in photoemission data is experimentally visible in an energy range down to 30 - 60 meV below the Fermi energy [32,33]. The shifts of spectral weight in the 2D iron-based superconductors are robust to 0.1 - 0.3 eV below the Fermi energy [28]. Therefore, it is essential to correct the electron concentration in superconducting state. We consider electron-phonon coupling in multi-energy layers, $\lambda_{PS} = 2\int \alpha_{PS}^2 \frac{F(\omega)}{\omega} d\omega$, where $\alpha_{PS} \sim \alpha_{E_F} C_F R_g$ [30]. The $R_g$ factor controls the amount of electrons below the Fermi level $E_F$ to participate in superconductivity, with the electron-phonon scattering matrix $g(E)$ in a state $E$ [27]. Suppose it is a phonon-mediated superconductor, the highest energy for excitation of electrons below $E_F$ cannot exceed the Debye energy $E_{Debye}$. Define

$$R_g = \frac{\left\langle \sum_{-\infty}^{E_F} g(E)\delta_A(E) \right\rangle}{\left\langle \sum_{-\infty}^{E_F} g(E)\delta_B(E) \right\rangle} \quad \text{where} \quad \delta_A(E) = 1 \text{ if } (E_F - E_{Debye}) \geq E \geq E_F. \text{ Similarly, } \delta_B(E) = 1$$

if $E = E_F$. Otherwise, $\delta_A(E) = \delta_B(E) = 0$. The electron-phonon scattering term on the Fermi surface is labeled as $\alpha_{E_F}$. The antiferromagnetic fluctuations $C_{AF}$ and the out-of-plane phonon $C_{ph}$ induced by the tetrahedral atoms amplify the electron-phonon scattering matrix terms by a factor of ~4 (abbreviated as Coh factor: $C_F = C_{AF}*C_{ph} = 4$) [24], where the value of Coh factor corresponds approximately to the antiferromagnetic amplification factor in NaFeAs and LiFeAs [25,26]. The out-of-plane vibration of Fe triggered by the tetrahedral atom induces electron's charges in xy-plane and the electron concentration across the tetrahedral bond is amended that induces the xy-potential [24]. With strong coupling, the electron-phonon coupling $\lambda_{PS}$ and the Coulomb pseudopotential $\mu$ are renormalized to $^*\lambda_{PS}$ and $\mu^*$ respectively [34]. Assuming that the pairing potential in the magnetic background corresponds to the first-order approximation, the pairing strength formula of iron-based superconductors in the presence of pressure $P$ is written as follows

$$\lambda = {}^*\lambda_{PS} f(E_{ex}) \quad \text{where} \quad f(E_{ex}) \sim \frac{[M_{Fe}M_{Fe}E_{co}]_{P>0}}{[M_{Fe}M_{Fe}E_{co}]_{P=0}} \quad [30]. \text{ The } M_{Fe} \text{ and } E_{co} \text{ are the}$$

magnetic moment of the Fe atoms and the exchange-correlation energy, respectively. The Debye temperature is acquired by $T_{Debye} = \frac{h}{2\pi k_B}\left[\frac{18\pi^2}{V}\frac{1}{\sum(1/v_s^3)}\right]^{1/3}$, where $h$, $k_B$, $V$, $v_s$ are the Planck constant, Boltzmann constant, the volume of the unit cell and the speed of sound [35].

The Debye temperature of the FeSe/SrTiO$_3$ is replaced by the vibrational energy of F-K phonon across the interface [29]. The DFT data is computed by WIEN2K, whereby the electronic properties are calculated by the GGA-PBE functional (unless otherwise specified) and the phonons are calculated by the Finite-Displacement Package [36-38]. The pairing strength is substituted into McMillian $T_c$ formula [39]. All Coulomb pseudopotentials are imported at 0.15, since the use of the conventional pseudopotential formula [34] to treat the highly correlated electron-electron interaction may not be accurate [40]. However, a pseudopotential within 0.1 to 0.2 is reasonable, because the error in the theoretical $T_c$ is only ~15% [30].

**Results**

Figure 1a shows that our pairing strength formula is applicable to the FeSe$_x$Te$_{1-x}$ system. The highest theoretical $T_c$ is located at $x = 0.25$ and the theoretical $T_c$ is reduced in the overdoped region. The decrease in the Debye temperature $T_{Debye}$ is observed when our calculated $T_{Debye}$ at $x = 0$, $x = 0.25$, $x = 0.5$, $x = 0.75$ are 240K, 195K, 180K and 120K, respectively. The pairing strength of FeSe is ~0.95 as shown in Figure 1b. The 25% doping of Te optimizes the pairing potential to ~0.99. Keep increasing the concentration of Te reduces the pairing strength. The pairing energies are almost identical at $x > 0.5$. In the absence of compression, all $f(E_{ex})$ equal one. Figure 2a demonstrates that our approach is not only valid at ambient pressure, but also successful in finite external pressure. Our model is more accurate when it calculates the $T_c$ of FeSe$_{0.5}$Te$_{0.5}$ at low pressure. Although the error in theoretical $T_c$ starts to increase above 4.5GPa, the theoretical $T_c$ distribution over the entire pressure range remains reasonable. Figure 2b confirms that the antiferromagnetic exchange interaction is inhibited under compression. The pairing strength becomes greatest at intermediate pressures. If the pressure exceeds 5GPa, the pairing strength is minimized.

Let's start our journey to acquire the theoretical $T_c$ of monolayer FeSe on a SrTiO$_3$ substrate step by step using the model of an antiferromagnetically-enhanced electron-phonon coupling. The flowchart is shown in Figure 3. After geometric optimization, the angles of the unit cell are 89.81º, 90.88º, 89.05º, with a tiny internal shear force being captured. The relaxed tetrahedral angle of Fe-Se-Fe is 108 degrees. The antiferromagnetic energy of FeSe can be amplified by low dimensionality when it is deposited in form of a monolayer on SrTiO$_3$ [23]. Compared to an FeSe monolayer without substrate, the FeSe film on SrTiO$_3$ shows an increased exchange correlation energy of ~16%. Apart from this, the local Fe moment in the isolated FeSe film is only ~0.5μ$_B$. However, the contact to SrTiO$_3$ amplifies the local Fe moment up to ~1.3μ$_B$. Our calculated the electron-phonon coupling on the Fermi surface without any amplification factor is $\lambda_{Fermi} = 0.12$. Based on our simulation, the antiferromagnetism of FeSe/SrTiO$_3$ is still as strong as of the FeSe monolayer without substrate. Hence the simultaneous occurrence of antiferromagnetism and tetrahedral atoms makes the Coh factor unavoidable [24]. The analytical result of C$_{AF}$ = 2 is used and our calculated C$_{Ph}$ in FeSe/SrTiO$_3$ is 2.9. After amplification of the Coh factor, the theoretical $T_c$ is only 14K. However, a massive enhancement of the pairing strength can be observed when the interfacial F-K phonon is involved [29]. The F-K phonon actuated via the interface contributes the vibrational energy of ~100meV (~1159K) [29]. With this enormous Debye temperature, the theoretical $T_c$ is increased to 69K, although the electron-phonon interaction is limited to the Fermi energy. In ARPES data it is evident that a shift of spectral weight occurs in the superconducting state 0.1~0.3eV below the Fermi level [28], which means that electrons in this energy range are affected by electron-phonon scattering as a result of the high phonon frequencies. This means that electrons in this energy range contribute to superconductivity, since the high phonon frequencies can scatter them up to the Fermi energy and need to be considered in the McMillan formula, and not only those at the Fermi energy as in the usual approximation applied to classical low-$T_c$ superconductors. The superconducting electron concentration is thus corrected and the average electron-phonon scattering matrix in these multi-energy layers is 1.96 times higher than the matrix considering only the Fermi level. This is the last factor with which our theoretical $T_c$ can reach 91K, which corresponds quite well to experimental $T_c$ of 100 K. All raw data used in our $T_c$ calculation of FeSe/SrTiO$_3$ are listed in the supplementary materials.

The pairing strength is renormalized as
$$^*\lambda_{PS} = \frac{\lambda_{PS}}{\lambda_{PS}+1} = \frac{R_g^2 C_F^2 \lambda_{Fermi}}{R_g^2 C_F^2 \lambda_{Fermi}+1} = \frac{(1.96^2)(2^2)(2.99^2)(0.12)}{(1.96^2)(2^2)(2.99^2)(0.12)+1} = 0.942$$

The pseudopotential is diluted as $\mu^* = \frac{\mu}{1+\lambda_{PS}} = \frac{0.15}{1+(1.96^2)(2^2)(2.99^2)(0.12)} = 0.0085$

We substitute all parameters into the McMillian $T_c$ formula,

$$T_c = \frac{T_{Debye}}{1.45}\exp\left(\frac{-1.04(1+{}^*\lambda_{PS})}{{}^*\lambda_{PS}-\mu^*(1+0.62{}^*\lambda_{PS})}\right) = \frac{1159}{1.45}\exp\left(\frac{-1.04(1+0.942)}{0.942-0.0085(1+0.62(0.942))}\right) = 91K$$

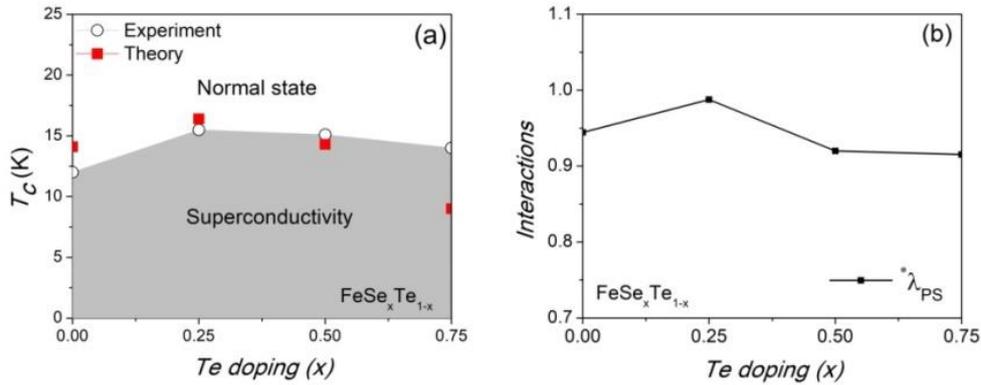

Figure 1: **a** Comparison between the theoretical and experimental $T_c$ of FeSe$_x$Te$_{1-x}$ [21]. **b** The pairing strength as a function of doping.

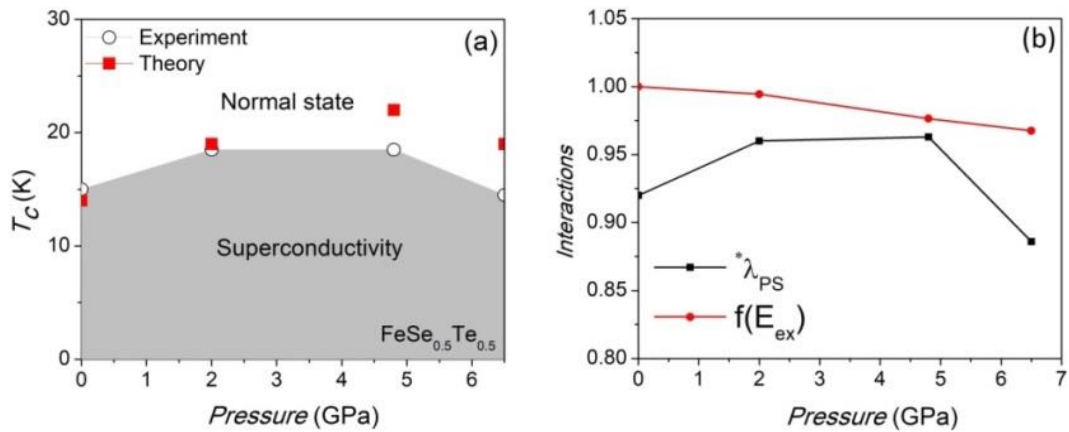

Figure 2: **a** The theoretical $T_c$ of the compressed FeSe$_{0.5}$Te$_{0.5}$ agrees with the experimental data [21]. **b** The individual interactions as a function of pressure.

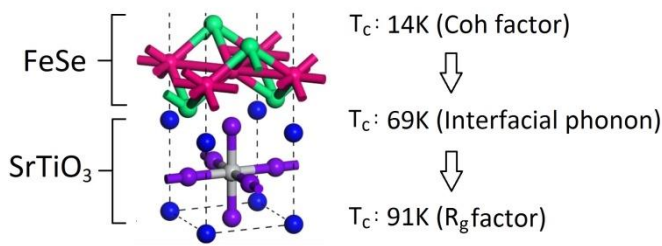

Figure 3: The local region of the unit cell. Our theoretical $T_c$ values after the amplifications of interfacial F-K phonon, Coh factor and $R_g$ factor [23].

**Discussion**

There are several effects that cause the $T_c$ of FeSe$_{0.25}$Te$_{0.75}$ to be the highest in Figure 1. Our simulation shows that $\lambda_{Fermi}$ at $x = 0.25$ is the largest, which strengthens the electron-phonon coupling at the Fermi surface. After considering the shift of spectral weight in the superconducting state well below $E_{Fermi}$ observed in ARPES data, the largest $R_g$ factor is also observed at $x = 0.25$, allowing a 4.9-fold increase in the average electron-phonon scattering matrix. Although the coupling strength at $x = 0.5$ and $x = 0.75$ is almost equal, the dramatic decrease of $T_{Debye}$ in the interval $0.5 < x < 0.75$ plays an important role in reducing the theoretical $T_c$ at $x > 0.5$. To investigate the pressure dependence of $T_c$ in FeSe$_{0.5}$Te$_{0.5}$, we compare $\lambda_{Fermi}$ and the $R_g$ factor. The variations of $\lambda_{Fermi}$ between 0.17 and 0.19 under pressure give only a tiny influence on the $T_c$ distribution. The control of the highest $T_c$ at intermediate pressure is mainly due to the $R_g$ factor. The $R_g$ factor of the uncompressed FeSe$_{0.5}$Te$_{0.5}$ is ~2, which is sufficient to form Cooper pairs at ~15K only. The external pressure of 2GPa and 4.8GPa increases the Debye temperature and presumably shifts the $R_g$ factor to 2.92 and 2.93, respectively. Despite the pressure beyond 5GPa strengthens the Debye energy even further [35], the excitation of the electrons becomes more difficult when the electron is too far below the Fermi level. This reduces the $R_g$ factor of FeSe$_{0.5}$Te$_{0.5}$ to 1.65 at 6.4GPa. After setting the pressure to 6.4GPa, the magnetic moment of Fe is reduced from 2.13$\mu_B$ to 2.06 $\mu_B$ as the orbital motion of the electrons is suppressed [35]. While the exchange correlation energy is only increased by 3% from 0GPa to 6.4GPa, the gain in exchange correlation energy cannot compensate for the loss of the magnetic moment and probably weakens the antiferromagnetism, as shown in Figure 2b.

The theoretical spring constant of FeSe bond is only 5 times smaller than the FeFe bond and therefore the out-of-plane vibration of Fe should exist. While Coh *et al* calibrated the GGA+A functional with experiment, they confirmed that the orthogonal phonon triggered the induced xy potential and reinforced the electron-phonon scattering matrix by a ratio of $C_{ph}$ = 2.2 [24]. It is possible to observe the induced xy potential at GGA level via the superposition principle where the upper tetrahedral plane '1' and the lower tetrahedral plane '2' are separately considered. We define $C_{ph} = \dfrac{0.5\left(qV_{ion}^{XY}DOS_1^{XY} + qV_{ion}^{XY}DOS_2^{XY}\right)}{qV_{ion}^{XY}DOS_{1\&2}^{XY}}$, where $DOS_c^{XY}$, $q$ and $V_{ion}^{XY}$ are the electronic density of states, Coulomb charge and the average ionic potential per atom in xy plane, respectively. The superposition principle guarantees the appearance of the orthogonal phonon even we do not calibrate the DFT functional. Our calculated C$_{ph}$ of the FeSe film is 2.9 which is comparable to their C$_{ph}$ value [24]. We test if the samples FeSe and FeSe$_x$Te$_{1-x}$ share the same value of C$_{ph}$. We choose x = 0.5 as an example. We have justified that our calculated C$_{ph}$ caused by the Se atom and Te atom are 3.31 and 1.32, respectively where the average C$_{ph}$ is 2.31. The C$_{ph}$ is nearly independent to pressure due to c >> a.

An empirical rule is that the $T_c$ of the iron-based superconductor is optimized when the tetrahedral angle is close to 109.5 degree [1]. When the FeSe monolayer is attached to the SrTiO$_3$, the tetrahedral angle is changed from 103 degrees to 108 degrees and the $T_c$ is benefits. However, all these antiferromagnetic and tetrahedral effects cannot explain the high $T_c$ near 100K until the interface properties are considered [29]. Despite the Debye temperature of the FeSe phonons (~250K) shows no significant size effect, an energetic F-K phonon carrying energy of 100meV (~1159K) was observed at the interface between the FeSe film and SrTiO$_3$ [29]. Since the 3D and 2D FeSe phonon are almost identical [29], the out-of-plane phonon from the tetrahedral sites should amplify the electron-phonon coupling of FeSe/SrTiO$_3$ by the same factor 2 [24]. Assuming that the F-K phonon and FeSe phonon interact with electrons simultaneously, two Debye energies, i.e. from the FeSe phonons and the F-K phonons, may influence the Cooper pairs. The two-fluid model, however, ensures that the onset $T_c$ is always related to the mechanism that gives the strongest pairing strength [42] and therefore choosing 1159K as the Debye temperature is justified.

The ARPES data of FeSe/SrTiO$_3$ show that the electrons in a wide range below the Fermi level ($\Delta E$ ~0.1 - 0.3eV) participate in superconductivity [28,29]. A question may be asked: Which energy source causes this shift of spectral weight? The F-K phonon may be one of the options since the $E_{Debye}$ is ~0.1eV [28,29]. Would it be exchange coupling? The exchange-correlation energy $E_{co}$ of FeSe/SrTiO$_3$ is also ~0.1-0.2eV. However, we believe that the F-K phonon is the energy source to generate this shift of spectral weight in FeSe/SrTiO$_3$. To support our argument, we revisit the ARPES results [32,33], where the bulk iron-based superconductors carrying $E_{co}$ ~ 0.1eV display a shift of spectral weight at $\Delta E$ ~ 30 - 60 meV below the Fermi level. If the shift is caused by the exchange-correlation energy, $\Delta E$ and $E_{co}$ should be comparable in the bulk iron-based superconductors, but this is not the case. If the exchange correlation energy is not the correct answer, we re-investigate the magnitude of $E_{Debye}$. Interestingly, the narrower range $\Delta E$ ~ 30 - 60 meV is comparable to the Debye temperature [44,45] of bulk iron-based superconductors. With this, we believe that $\Delta E$ ~ $E_{Debye}$ is unlikely to be a coincidence. The shift of spectral weight in ARPES in iron-based superconductors is thus likely triggered by phonon-mediated processes. After revising the electron concentration in the superconducting state, our calculated $T_c$ is further increased to 91K. We have verified that the Coh factor is only reduced by ~3% at $E_F$ - 100meV.

On the Fermi surface, a nematic order is observed in various iron-based superconductors [1,19,43] and the electron-electron interaction should be influenced accordingly. Although our approach does not consider the nematic order, our approach averages the electron-phonon coupling between $E_F$ - $E_{Debye}$ and $E_F$, which minimizes the error due to the nematic order at the Fermi surface. From a mathematical point of view, the $\alpha_{PS}$ is calculated by $\alpha_{E_F} C_F R_g$, where the Coh factor $C_F$ is a constant. The $\alpha_{E_F}$ is directly proportional to the $|g(E_F)|$. If the nematic order changes the $g(E_F)$ value, the $R_g$ factor cancels the nematic contribution because the $R_g$ is inversely proportional to $|g(E_F)|$. The numerator of $R_g$ contains the average electron-phonon scattering matrix in multi-energy layers, where the Fermi energy is only one of them. Under these circumstances, the error of $\alpha_{PS}$ from neglecting the nematic effect is relatively small and our $T_c$ calculation should remain accurate.

**Summary**

We have presented a model that considers a combination of electron-phonon coupling and antiferromagnetic fluctuations as a possible method to accurately calculate the $T_c$ of iron-based '11- type' superconductors, including their pressure and doping dependence. When applied to monolayer FeSe on a SrTiO$_3$ substrate, we find that the interfacial phonons are of major importance to explain the high temperature superconductivity.

**References**

[1] S Fujitsu, S Matsuishi & H Hosono To, Iron based superconductors processing and properties, *International Materials Reviews* **57**, 311-327 (2012).

[2] G. R. Stewart, Superconductivity in iron compounds, *Rev. Mod. Phys.* **83** (2011) 1589-1652.

[3] P. J. Hirschfeld, M. M. Korshunov and I. I. Mazin, Gap symmetry and structure of Fe-based superconductors, *Rep. Prog. Phys.* **74** (2011) 124508.

[4] V. Cvetkovic and Z. Tesanovic, Valley density-wave and multiband superconductivity in iron-based pnictide superconductors, *Phys. Rev. B* **80** (2009) 024512.

[5] I. I. Mazin and J. Schmalian, Pairing symmetry and pairing state in ferropnictides: Theoretical overview, *Physica C* **469** (2009) 614-627.

[6] V. Cvetkovic and Z. Tesanovic, Multiband magnetism and superconductivity in Fe-based compounds, *Europhys. Lett.* **85** (2016) 37002.

[7] R. M. Fernandes, D. K. Paratt, W. Tian, J. Zarestky, A. Kreyssig, S. Nandi and M. G. Kim, Unconventional pairing in the iron arsenide superconductors, *Phys. Rev. B* **81** (2010) 140501(R).

[8] For a review see: A. V. Chubukov and P. J. Hirschfeld, Iron-based superconductors, seven years later, *Physics Today* **68** (2015) 46-52

[9] J.-H. Chu, J. G. Analytis, K. De Greve, P. L. McMahon, Z. Islam, Y. Yamamoto and I. R. Fisher, In-plane resistivity anisotropy in an underdoped iron arsenide superconductor, *Science* **329** (2010) 824-826.

[10] S. Jiang, H. S. Jeevan, J. Dong and P. Gegenwart, Thermopower as a Sensitive Probe of Electronic Nematicity in Iron Pnictides, *Phys. Rev. Lett.* **110** (2013) 067001.

[11] A. Dusza, A. Lucarelli, F. Pfuner, J.-H. Chu, I. R. Fisher and L. Degiorgi, Anisotropic charge dynamics in detwinned Ba(Fe1-xCox)2As2, *Europhys. Lett.* **93** (2011) 37002.

[12] S. Kasahara, H. J. Shi, K. Hashimoto, S. Tonegawa, Y. Mizukami, T. Shibauchi, K. Sugimoto, T. Fukuda, T. Terashima, A. H. Nevidomskyy and Y. Matsuda, Electronic nematicity above the structural and superconducting transition in $BaFe_2(As_{1-x}P_x)_2$, *Nature (London)* **486** (2012) 382-385.

[13] M. Fu, D. A. Torchetti, T. Imai, F. L. Ning, J.-Q. Yan and A. S. Sefat, NMR Search for the Spin Nematic State in a LaFeAsO Single Crystal, *Phys. Rev. Lett.* **109** (2012) 247001.

[14] T.-M. Chuang, M. P. Allan, J. Lee, Y. Xie, N. Ni, S. L. Bud'ko, G. S. Boebinger, P. C. Canfield and J. C. Davis, Nematic Electronic Structure in the "Parent" State of the Iron-Based Superconductor $Ca(Fe_{1-x}Co_x)_2As_2$, *Science* **327** (2010) 181-184.

[15] R. M. Fernandes, A. V. Chubukov and J. Schmalian, What drives nematic order in iron-based superconductors?, *Nat. Phys.* **10** (2014) 97-104.

[16] R. M. Fernandes, P. P. Orth, J. Schmalian, Intertwined Vestigial Order in Quantum Materials: Nematicity and Beyond, *Annu. Rev. Condens. Matter Phys.* **10**:133–5(2019) and references therein.

[17] F.C. Hsu, J.Y. Luo, K.W. Yeh, T. K Chen, T.W Huang, Phillip M. Wu, Y.C. Lee, Y.L. Huang, Y.Y. Chu, D.C. Yan, and M.K. Wu, Superconductivity in the PbO-type structure α-FeSe. *Proc. Natl. Acad. Sci. USA*. **105**, 14262–14264 (2008).

[18] A. Böhmer and A. Kreisel, Nematicity, magnetism and superconductivity in FeSe. *J. Phys. Condens. Matter* **30**, 023001 (2018), and references therein.

[19] J. Kang, R. M. Fernandes, and A. Chubukov, Superconductivity in FeSe: The Role of Nematic Order, *Phys. Rev. Lett.* **120**, 267001 (2018).

[20] S. Medvedev, T. M. McQueen, I. A. Troyan, T. Palasyuk, M. I. Eremets, R. J. Cava, S. Naghavi, F. Casper, V. Ksenofontov, G. Wortmann, C, Felser., Electronic and Magnetic Phase Diagram of β-$Fe_{1.01}Se$ with superconductivity at 36.7 K under pressure. *Nature Materials*. **8**, 630–633 (2009).

[21] H. Takahashi, T. Tomita, H. Takahashi, Y. Mizuguchi, Y.Takano, S. Nakano, K. Matsubayashi & Y. Uwatoko, High-pressure studies on $T_c$ and crystal structure of iron chalcogenide superconductors, *Sci. Technol. Adv. Mater.* **13**, 054401 (2012).

[22] R. Peng, X. P. Shen, X. Xie, H. C. Xu, S. Y. Tan, M. Xia, T. Zhang, H. Y. Cao, X. G. Gong, J. P. Hu, B. P. Xie, D. L. Feng, Enhanced superconductivity and evidence for novel pairing in single-layer FeSe on SrTiO3 thin film under large tensile strain. *Phys. Rev. Lett.* **112**, 107001 (2014).

[23] J.F. Ge, Z.L. Liu, C. Liu, C.L. Gao, D. Qian, Q.K. Xue, Y. Liu & J.F. Jia, Superconductivity above 100 K in single-layer FeSe films on doped $SrTiO_3$, *Nature Materials,* **14**, 285–289 (2015).

[24] S. Coh, M. L. Cohen, S. G. Louie, Antiferromagnetism enables electron-phonon coupling in iron-based superconductors, *Phys. Rev. B* **94**, 104505 (2016).

[25] B. Li, Z. W. Xing, G. Q. Huang, M. Liu, Magnetic-enhanced electron-phonon coupling and vacancy effect in "111"-type iron pnictides from first-principle calculations, *J. App. Phys.* **111**, 033922 (2012).

[26] S. Deng, J. Köhler, A. Simon, Electronic structure and lattice dynamics of NaFeAs, *Phys. Rev. B* **80**, 214508 (2009).

[27] Z. Wang, C. Liu, Y. Liu and J. Wang, High-temperature superconductivity in one-unit-cell FeSe films, *J. Phys.: Cond. Mat.* **29**, 153001 (2017).

[28] C. Zhang, Z. Liu, Z. Chen, Y. Xie, et al, Ubiquitous strong electron–phonon coupling at the interface of $FeSe/SrTiO_3$, *Nat. Commun.* **8**:14468 (2017).

[29] S. Zhang, J. Guan, Y.Wang, T. Berlijn, et al, Lattice dynamics of ultrathin FeSe films on $SrTiO_3$, *Phys. Rev. B* **97**, 035408 (2018).

[30] C. H. Wong and R. Lortz, The antiferromagnetic and phonon-mediated model of the NaFeAs, LiFeAs and FeSe superconductors, arXiv:1902.06463.

[31] C. H. Wong and R. Lortz, to be published.

[32] X.-W. Jia et al., Common Features in Electronic Structure of the Oxypnictide Superconductor from Photoemission Spectroscopy, *Chinese Phys. Lett.* **25**, 3765-3768 (2008).

[33] U. Stockert, M. Abdel-Hafiez, D. V. Evtushinsky, V. B. Zabolotnyy, A. U. B. Wolter, S. Wurmehl, I. Morozov, R. Klingeler, S. V. Borisenko, B. Büchner, Specific heat and angle-resolved photoemission spectroscopy study of the superconducting gaps in LiFeAs, *Phys. Rev. B* **83**, 224512 (2011).

[34] K.C Weng, C. D. Hu, The p-wave superconductivity in the presence of Rashba interaction in 2DEG, *Sci. Rep.* **6**, 29919 (2016).

[35] J. R. Christman, Fundamentals of solid state physics (Wiley, 1988).

[36] P. Blaha, K. Schwarz, G. K. H. Madsen, D. Kvasnicka and J. Luitz, WIEN2k, An Augmented Plane Wave + Local Orbitals Program for Calculating Crystal Properties (Karlheinz Schwarz, Techn. Universität Wien, Austria) (2001).

[37] J. P. Perdew, J. A. Chevary, S. H. Vosko, K. A. Jackson, M. R. Pederson, D. J. Singh, C. Fiolhais, Atoms, molecules, solids, and surfaces: Applications of the generalized gradient approximation for exchange and correlation, *Phys. Rev. B* **46**, 6671 (1992).

[38] A. D. Becke, Density-functional exchange-energy approximation with correct asymptotic behavior, *Phys. Rev. A* **38**, 3098 (1988).

[39] W. L. McMillian, Transition Temperature of Strong-Coupled Superconductors, *Phys. Rev.* **167**, 331 (1968).

[40] E. J. König, P. Coleman, The Coulomb problem in iron based superconductors, arXiv:1802.10580 (2018).

[41] S. Coh, M. L. Cohen and S.G. Louie, Large electron–phonon interactions from FeSe phonons in a monolayer, *New J. Phys.* **17** (2015) 073027.

[42] M. Tinkham, Introduction to superconductivity (Dover Publications, 1996).

[43] T. Hanaguri, K. Iwaya, Y. Kohsaka, T. Machida, T. Watashige, S. Kasahara, T. Shibauchi and Y. Matsuda, Two distinct superconducting pairing states divided by the nematic end point in FeSe$_{1-x}$S$_x$, *Sci. Adv.* 4(5):eaar6419 (2018).

[44] R. A. Shukor, Calculated Sound Velocity Change in LaFeAsO$_{0.89}$F$_{0.11}$ at the Superconducting Transition, *J. Supercond. Nov. Magn.* **23**: 1229–1230 (2010).

[45] Y. Wen, D. Wu, R. Cao, L. Liu, L. Song, The Third-Order Elastic Moduli and Debye Temperature of SrFe$_2$As$_2$ and BaFe$_2$As$_2$: a First-Principles Study, *J. Supercond. Nov. Magn.* **30**, 1749–1756 (2017).

Supplementary materials

Bulk FeSe$_x$Te$_{1-x}$

| x | a (Å) | c (Å) | $\lambda_{Fermi}$ | $R_g$ | Debye (K) |
|---|---|---|---|---|---|
| 0 | 3.7676 | 5.4847 | 0.12 | 3.04 | 240 |
| 0.25 | 3.8129 | 6.1500 | 0.21 | 4.92 | 195 |
| 0.5 | 3.8003 | 5.9540 | 0.18 | 2.02 | 180 |
| 0.75 | 3.7872 | 5.6492 | 0.17 | 1.99 | 120 |

Bulk FeSe$_{0.5}$Te$_{0.5}$

| P(GPa) | a (Å) | c (Å) | $\lambda_{Fermi}$ | $R_g$ | Debye (K) |
|---|---|---|---|---|---|
| 0 | 3.8003 | 5.9540 | 0.18 | 2.02 | 190 |
| 2 | 3.7760 | 5.8623 | 0.17 | 2.92 | 230 |
| 4.8 | 3.7425 | 5.7352 | 0.19 | 2.93 | 280 |
| 6.4 | 3.7229 | 5.6068 | 0.17 | 1.65 | 290 |

FeSe/SrTiO$_3$

| a (Å) | b (Å) | c (Å) | D (Å) | $\lambda_{Fermi}$ | $R_g$ | Debye (K) |
|---|---|---|---|---|---|---|
| 3.8197 | 3.8698 | 5.9540 | 52.484Å | 1.6 | 1.96 | 1159 |

[*]The unit cell of FeSe/SrTiO$_3$ occupied the volume of 3.8197 Å x 3.8698 Å x 5.9540 Å. The layer-to-layer distance *D* is 52.484Å